\documentclass[twocolumn,a4paper,superscriptaddress,floatfix,longbibliography]{revtex4}
\usepackage{graphicx,amsfonts,amsmath,bm,color,tabularx}
\usepackage{multirow}
\usepackage[normalem]{ulem}

\begin{document}

\title{Low-dimensional Heisenberg magnets: Riemann zeta function regularization}

\author{Valentin Yu. Irkhin}
\affiliation{M.N. Mikheev Institute of Metal Physics, S. Kovalevskaya str. 18, 620108 Ekaterinburg, Russia}

\date{\today}

\date{\today}
\begin{abstract}
The Riemann zeta function regularization is employed to extract finite temperature corrections to effective magnetic moment $S^*$ of one- and two-dimensional Heisenberg ferro- and antiferromagnets. Whereas for the one-dimensional ferromagnet we obtain the usual $T^{1/2}$ spin-wave dependence, for  the antiferromagnetic chain the  dependence is described by a generalized incomplete Riemann  function.
The quantity  $S^*$ determines strong short-range magnetic order  in the absence of long-range order, in particular the correlation length.
For the one-dimensional ferromagnet, the results are confirmed by the self-consistent spin-wave theory and Monte Carlo simulations  by Takahashi \textit{et al}.
\end{abstract}

\maketitle

\section{Introduction}

Riemann's zeta function 
\begin{equation}
\zeta(x>1)=\sum_{n=1}^{\infty} n^{-x} =\frac{1}{\Gamma (x)}\int_0^\infty \frac{z^{x-1}dz}{\exp z -1}
\end{equation}
(where $\Gamma (x)$ is the Euler Gamma function), 
being one of most important function in mathematics, finds also applications in physics \cite{Elizalde,Elizalde1,Schumayer}. Especially this  concerns quantum field theory where regularization of divergences is achieved by assigning finite values to observables. The regularization is achieved by analytical continuation of  (1) to any $x<1$ and into the complex plane, which enables one to sum up divergent series.

A well-known example of the zeta-regularization is description of the Casimir effect, i.e., occurrence of a physical force acting on the macroscopic boundaries of a confined space, which arises from the quantum fluctuations of electromagnetic field \cite{Casimir,Casimir1}. A finite value of the Casimir force is obtained by using  $\zeta(-3)=-1/120$. Recently, the zeta-regularization has been exploited to construct an effective action for magnetoelectric effect in multilayer graphene \cite{KG}.
Here we demonstrate that a similar approach can be applied in the quantum magnetism theory.

According to the Mermin-Wagner theorem, isotropic one- and two-dimensional (1d and 2d) systems cannot have long-range order at finite temperatures $T$ because of strong fluctuations, contribution of which being divergent. { 
Such a situation occurs in the isotropic Heisenberg model where long-range order in 1d and 2d cases does not exist at finite $T$, and for 1d antiferromagnets even in the ground state. This is in contrast with the anisotropic 2d Ising model where a true long-range order is retained at finite $T$ (in the 2d XY-model, a quasi-long-range order occurs below the Berezinskii--Kosterlitz--Thouless transition temperature). 

At the same time, low-dimensional Heisenberg systems exhibit considerable short-range order at low $T$: the correlation length becomes very large, and the system is nearly ordered.
Describing the thermodynamics of such systems is the aim of the present paper. In particular, we show that the short-range order parameter of the one-dimensional Heisenberg ferromagnet demonstrates spin-wave temperature dependence similar to that of the long-range order parameter in higher-dimensional systems.}

Let us write down the magnon correction to magnetization of a Heisenberg ferromagnet on the $d$-dimensional cubic lattice with the lattice constant $a=1$:
\begin{equation}
	\delta \overline S =-\sum_{\bf q} N_{\bf q}=-\frac{1}{(2\pi)^d}\int_0^\infty \frac{d^{d}q}{\exp(Dq^2/T)-1}
    \label{S}
\end{equation}
where $N_{\bf q}=\langle b^\dagger_{\bf q} b_{\bf q}\rangle$, $b^\dagger_{\bf q},\, b_{\bf q}$ are magnon operators, $D$  is the spin-wave stiffness, $d^{d}q= 2\pi^{d/2} q^{d-1}dq/\Gamma(d/2)$. 
 For $d>2$ this integral is convergent and is calculated as 
 \begin{equation}
	\delta \overline S =-\frac{\zeta (d/2)}{(4 \pi)^{d/2}} \left( \frac{T}{D} \right)^{d/2}\
    \label{S1}.
\end{equation}
For $d \le 2$ a divergence occurs, but we can try to assign a finite value
 to the integral by using the values of $\zeta(x)$ obtained by analytical continuation.
Then for $d=1$  we have
\begin{equation}
	\delta \overline S =-\frac{\zeta (1/2)}{2 \pi^{1/2}} \left( \frac{T}{D} \right)^{1/2}
    \label{S2}
\end{equation}
with $\zeta(1/2)=-1.460$.
To clarify physical sense of this correction and relate it to observables we will perform below a treatment within the formalism of the self-consistent spin-wave theory (SSWT) \cite{Takahashi,Takahashi1}. We will demonstrate that in fact this describes characteristics of strong short-range order in the absence of long-range order.

Unfortunately, for $d=2$ we meet the pole of the function $\zeta(x)$ at $x=1$,  so that a finite value is not obtained. We will see below that the leading order correction is absent, but we can extract finite values from next-order term in the integral expansion. For an antiferromagnet, the situation  turns out to be still more complicated and is resolved by using a truncated (incomplete) Riemann function.

\section{Self-consistent spin-wave theory}
Spin-wave theory does describe 2d and even 1d systems  \cite{Takahashi1} provided that we use a self-consistent version.
We start from the Heisenberg Hamiltonian
\begin{equation}
	H=\sum_{ij}J_{ij}\mathbf{S}_{i}%
	\mathbf{S}_{j} \label{H}.
\end{equation}

\subsection{Ferromagnets}
The self-consistency equation of SSWT for a ferromagnet is condition of zero magnetization, $\overline S=0$, i.e., 
\begin{equation}
	 S =\sum_{\bf q} \langle b^\dagger_{\bf q} b_{\bf q}\rangle  =\frac{1}{(2\pi)^d}\int_0^\infty \frac{d^{d}q}{\exp([Dq^2-\mu]/T)-1}
         \label{H1}.
\end{equation}
where $\mu$ is the chemical potential of magnons which is introduced to ensure the finite values of magnons at a site.

The expansion of the Bose-Einstein integrals in $y=-\mu/T$ by using the contour integration in the complex plane has the form  \cite{Robinson}
\begin{eqnarray}
	F(x,y)=\frac{1}{\Gamma (x) } \int_0^\infty \frac{q^{x-1}dq}{\exp(q-y)-1} \nonumber \\
    =\Gamma (1-x) y^{x-1}+\sum_{n=0}^{\infty}(-y)^n\zeta(x-n)
/n!    \label{mu}
\end{eqnarray}
 where $x$ is assumed to be non-integer.
For positive integer $x=m$, infinite terms occur in the expansion, but the singularities can be canceled to obtain \cite{Robinson}
\begin{eqnarray} 
F(m,y)=\{C&+&\Gamma' (m)/\Gamma (m)-\ln y\} (-y)^{m-1}/\Gamma (m) \nonumber\\
&+&\sum_{n\neq m-1}^{\infty}(-y)^n\zeta(m-n)
/n!    
\label{mu2}
\end{eqnarray}
where $C=0.577$ is the Euler-Mascheroni constant.
For $m=1$, $C=-\Gamma' (m)/\Gamma (m)$,  and the singular $\zeta(1)$ term does not work.
Then we have
\begin{equation}
	\sum_{\bf q} N_{\bf q}=\Gamma (d/2)F(d/2,-\mu/T)
    \label{chi1}.
\end{equation}
The spin-wave magnetic susceptibility is determined by spin correlation function \cite{Takahashi1}:
\begin{eqnarray}
    	\chi= \frac{1} {T} \sum_i \langle S^z_0 S^z_i\rangle  = \frac{1}{3T} \sum_{\bf q} N_{\bf q}(1+ N_{\bf q}) \nonumber \\
        =\frac{1}{3T} \Gamma (d/2)F(d/2-1,-\mu/T)
    \label{chi}.
\end{eqnarray}

For 1d ferromagnet the expansion of the equation (\ref{H1}) according to (\ref{mu})  yields
\begin{equation}
	S= \frac{T} {2\mu}+\frac{ \zeta(1/2)} {2 \pi^{1/2}} \left( \frac{T}{D} \right)^{1/2} +...  \label{w}.
\end{equation}
It is important to note that we do not take into account here  the renormalization of the spin-wave stiffness $D=|dJS|$  since for $d=1$ it contains higher powers of temperature ($\delta D(T)\propto T^{d/2+1} $) and can be neglected.
Solving this we derive { the low-temperature asymptotics}
\begin{equation}
	|\mu/T|^{-1/2}= 2S^*, \, S^*=S+\delta \overline S \label{ww}
\end{equation}
with $\delta \overline S$ being just the spin-wave correction determined by (\ref{S2}).
{ 
According to Ref.\cite{Takahashi1}, the correlation length in SSWT is determined by the small-$q$ expansion
\begin{equation}
	N_{\bf q}=\frac{1} {\exp [(Dq^2-\mu)/T]-1}\simeq\frac{Td/D} {q^2+(2\xi)^{-2}}.
\end{equation}
 This definition with $2\xi$ in denominator is justified by that, with account of the constraint (\ref{H1}), the spin correlation function in SSWT takes the form
\begin{equation} 
	\langle {\bf S}_i {\bf S}_j\rangle =  \left( \sum_{\bf q} \cos [{\bf q} ({\bf R}_i-{\bf R}_j)]N_{\bf q} \right)^2 \propto e^{-|{\bf R}_i-{\bf R}_j|/\xi}.
    \end{equation}
Then the correlation length is related to the chemical potential  by 
\begin{equation}
	\xi=  \frac{1} {2} |D/d\mu| ^{1/2}.
    \label{xi}
\end{equation}}
For $d=1$ we have $\xi\simeq S^*D/T$, and the susceptibility is obtained from (\ref{chi}) as
\begin{equation}
	\chi= \frac{D} {12T}|\mu/T|^{-3/2}=\frac{2D(S^*)^3} {3T^2}
    \label{chi11}.
\end{equation}
According to Takahashi \cite{Takahashi1}, who obtained (without detailed physical interpretation) the series expansion of (\ref{chi11}), this result in an excellent agreement with the Monte Carlo simulations for $S=1/2$ \cite{Takahashi2}.

For a 2d ferromagnet we have from (\ref{mu2})
\begin{equation}
	|\mu|/T= \exp  \left( -\frac{2 \pi DS^*}{T} \right) 
         \end{equation}
so that the correlation length and magnetic susceptibility
\begin{equation}
	\chi=\frac{2}{3\pi D} \exp  \left( \frac{2 \pi DS^*}{T} \right). 
         \end{equation}
are exponentially large \cite{Takahashi,IK1,Kopietz}.

The lowest-order temperature corrections to $S^*$ are absent. 
To obtain next order correction we can, following to Ref. \cite{Takahashi}, use the expansion of the magnon density of states
$$ \delta g(\omega) =g(\omega)-\frac{1}{4\pi} = \frac{\omega}{32 \pi}.$$
The resulting integral is not singular at $\mu =0$ and we obtain the spin-wave correction
\begin{equation}
	\delta \overline S =-\int_0^\infty \frac{\delta g(\omega)d\omega}{\exp(\omega/T)-1} =-\frac{\zeta (2)}{8 \pi} \left( \frac{T}{D} \right)^{2}.
\end{equation}
Thus $S^*$ has the same $T^2$ dependence as  $D$.

We can also consider a somewhat exotic case of frustrated ferromagnet where the spin stiffness tends to zero owing to competing exchange interactions \cite{Ign}, $\omega_{{\bf q }\rightarrow{0}} \simeq B q^4$.
Then the long-range order is absent for finite $T$, but we obtain a finite correction to the short-range order parameter
\begin{equation}
	\delta \overline S =-\frac{\zeta(d/4) \Gamma(d/4)}{2(4 \pi)^{d/2}\Gamma(d/2)} 
    \left( \frac{T}{B} \right)^{d/4}
    \label{S21}.
\end{equation}

\subsection{Antiferromagnets}

The situation for an antiferromagnet differs by the presence of zero-point oscillations. The SSWT equation has the form
\begin{equation}
	 S+ \frac {1} {2} =\sum_{\bf q} \frac {2SJ_0-\mu}{2\omega_{\bf q}} (1+2N_{\bf q})
              \label{a1}.
\end{equation}
where 
\begin{eqnarray}
  N_{\bf q}&=&\frac{1}{\exp (\omega_{\bf q}/T)-1}, \nonumber \\ 
  \omega_{\bf q} &=& [(2SJ_0-\mu)^2-4S^2 J_{\bf q}^2]^{1/2}, \nonumber\\
  \omega_{{\bf q, |q-Q|} \rightarrow {0}} &\simeq& (c^2q^2+ 4|\mu|J_0S)  ^{1/2},
\end{eqnarray}
$c\simeq 2SJ_0d^{1/2}$ being the spin-wave velocity, $|\mu| \ll c$, ${\bf Q}$ is the antiferromagnetic wavevector. Again, $c$ is somewhat renormalized by standard  zero-point and high order temperature corrections ($\delta c(T)\propto T^{d+1}$), but we will not discuss them for brevity.

For $d=1$ zero-point oscillations result in vanishing of the ground state sublattice magnetization. For half-integer spins the magnon spectrum is gapless, but for integer spins   the Haldane gap $\xi^{-1} \propto \Delta \propto e^{-\pi S}$ occurs in the system \cite{Haldane}.
For integer  $S$, SSWT can provide a rough qualitative description \cite{Yamamoto}. We obtain from (\ref{a1})
\begin{equation}
S +\frac {1} { 2}=\frac {1} { 2\pi} \ln  \frac {c} {|\mu|}+\frac {2T} { \pi c} \tilde \zeta(0,|\mu c|^{1/2}/T)
              \label{a11}
\end{equation}
where
\begin{equation}
\tilde \zeta(x,y)=\int_y^\infty \frac{(z^2-y^2)^{(x-1)/2}dz}{\exp z -1}
\label{rim}
\end{equation}
is a kind of generalized (upper incomplete) Riemann's  function, $\tilde \zeta(0,y \rightarrow{0}) \simeq {\pi}/(2y)$. 
Using this function enables one to regularize the integrals since they become finite for $x=0,\,1$ at finite $y$.
Note that similar functions occur in the theory of degenerate Bose gas  \cite{Landau} -- a problem which is similar to the Haldane chain description \cite{Jolicoeur}.

Now $\mu = \mu (T,S)$  determined as the solution to the equation
(\ref{a11})  is finite at $T=0$ owing to the Haldane gap, $\mu (0,S) \propto \xi^{-2} \propto \Delta^2$. 
For $T \gg \Delta$ we return to usual spin-wave description.


The static susceptibility at low temperatures $c \gg T \gg \Delta$ reads
\begin{equation}
	\chi= \frac{4T}{3 c}|2\mu c|^{-1/2} 
    \label{chi1}
\end{equation}
and is exponentially small at $T \ll  \Delta$.
The staggered susceptibility is   considerably enhanced:
\begin{equation}
	\chi_Q =\frac{1} {T} \sum_i \langle S^z_0 S^z_i\rangle  \exp(i\pi R_i)= \frac{T}{3\pi|\mu| c} 
    \label{chi2}.
\end{equation}
This simple interpolation description of the Haldane chain differs  from more rigorous results obtained by a rescaling in terms of the energy gap \cite{Jolicoeur}. Besides that, modified spin-wave theories cannot provide quantitatively the values of  $\Delta$, provided by Monte Carlo simulations \cite{Yamamoto} (this quantity is not determined in the approach \cite{Jolicoeur}).

For $d=2$   the ground state sublattice magnetization 
\begin{equation}
	 S_0 =S- \frac{1}{2}\sum_{\bf q} \left( \frac {2SJ_0-\mu}{2\omega_{\bf q}} -1 \right) \simeq S- 0.1966
         \end{equation}
is finite. 
The equation (\ref{a1}) takes the form
\begin{equation}
	 S_0 =\frac {T} {\sqrt 2 \pi c} \tilde \zeta(1,|\mu c \sqrt 2 |^{1/2}/T).
              \end{equation} { 
For $x=1$, the function (\ref{rim}) is reduced to an elementary integral containing the exponential function:
\begin{equation}
	 \tilde \zeta(1,y) = y-\ln(e^{y}-1)=-\ln y+y/2+O(y^2).
     \label{rim1}
              \end{equation}
Retaining the $\ln y$ term in (\ref{rim1}) only, we derive
\begin{equation}
	\mu/c= -\sqrt 2\xi^{-2}/8 = - 2^{-3/2} \left(\frac{ T}{c} \right)^2 \exp  \left( -\frac{2^{3/2} \pi cS^*}{T} \right). 
         \end{equation}     
Corrections owing to the $O(y)$ terms in (\ref{rim1}) can be neglected since $y \sim |\mu|^{1/2}/T$ is exponentially small.}
         
Then the staggered susceptibility and corresponding correlation function are exponentially large \cite{Chakravarty,Takahashi,IK}:
\begin{equation}
	\chi_{\bf Q} = \frac{1}{T}\langle S^z_{\bf Q}S^z_{\bf -Q}\rangle 
    \simeq \frac{2}{\pi} \exp  \left( \frac{2^{3/2}\pi cS^*}{T} \right),
             \end{equation}           
whereas the static susceptibility is nearly constant, $\chi \simeq 2^{1/2}S^*/ (12 c) +O(T)$.
The short-range order is very strong: the spin spectral density behaves as 
\begin{equation}K({\bf q}, \omega) \simeq S^{*2}\delta({\bf q-Q},\xi)\delta(\omega,\xi), 
\end{equation} 
where the delta-functions are smeared on the scales $1/\xi$ and $J/\xi$, respectively \cite{IK}. 
The temperature correction  in $S^*$ turns out to be proportional to $T^3$ \cite{Takahashi}, as well as the correction to $c$.

For frustrated  antiferromagnets \cite{IK2} we have $c\rightarrow{0}$, $\omega_{{\bf q} \rightarrow {0}} \simeq (b^2q^4+ 4|\mu|J_0S)  ^{1/2}$. In the 2d case we can obtain the equation for the chemical potential, which is similar to (\ref{a11}),
resulting in the ground state spin-wave gap.
As well as for a  frustrated ferromagnet,  corrections $\delta \overline S$ for a  frustrated 3d antiferromagnet can be derived  in terms of a generalized zeta function.


\section{Conclusions}

We have demonstrated that using the Riemann zeta function and related functions can be useful in the magnetism theory. The corresponding regularization enables one to calculate  temperature  spin-wave corrections to short-range parameters, especially useful results being obtained in the 1d case.

It is interesting that we have two sets of short-range order parameters: characteristics of spin-wave spectrum $D$, $c$ and parameters $\mu$, $S^*$, the corresponding temperature dependences being different for $d=1$.  All these parameters enter observable quantities.
The  effective magnetic moment $S^*$ can be observed, e.g., as a quasi-splitting for the spectrum of an electron interacting with localized spins, see calculations in the $s-d$ exchange model \cite{IK}.

When calculating thermodynamic and transport properties, the zeta-function regularization may provide an alternative method to the renormalization group approach in the case of power-law rather than logarithmic divergences.
A similar procedure can be performed with Fermi-Dirac integrals, which occur for low-dimensional systems, e.g., in the presence of strong van Hove singularities. 

The author is grateful to  M.I. Katsnelson for useful discussion. 
 The investigation was supported by state assignment of Ministry of Science and Higher Education of the Russian Federation for the Institute of Metal Physics.

\end{document}